\documentclass[aps,prd,nofootinbib,twocolumn]{revtex4}
\usepackage{amssymb}
\usepackage{amsmath,color}
\usepackage{epsfig}
\usepackage{graphicx,epsf,epsfig}
\usepackage{bbm}
\usepackage{subfigure}
\usepackage{multirow}
\usepackage{setspace}
\usepackage{verbatim}
\usepackage{epstopdf}
\usepackage{ulem}
\usepackage[linktocpage]{hyperref}
\usepackage{multirow}
\usepackage[marginal]{footmisc}

\begin{document}
\title{Dependence of the amplitude of gravitational waves from preheating on the inflationary energy scale}
\author{Rong-Gen Cai$^{1,2,3}$}
\email{cairg@itp.ac.cn}

\author{Pei-Ze Ding$^{4}$}
\email{peize18@mail.ustc.edu.cn}

\author{Zong-Kuan Guo$^{1,2,3}$}
\email{guozk@itp.ac.cn}

\author{Chengjie Fu$^{3}$}
\email{fucj@itp.ac.cn}

\author{Jing Liu$^{1,2}$}
\email[Corresponding author:]{liujing@ucas.ac.cn}

\affiliation{$^{1}$School of Fundamental Physics and Mathematical Sciences, Hangzhou Institute for Advanced Study, University of Chinese Academy of Sciences, Hangzhou 310024, China
}

\affiliation{$^{2}$School of Physical Sciences, University of Chinese Academy of Sciences,
	No.19A Yuquan Road, Beijing 100049, China}

\affiliation{$^{3}$CAS Key Laboratory of Theoretical Physics, Institute of Theoretical Physics,
	Chinese Academy of Sciences, P.O. Box 2735, Beijing 100190, China}

\affiliation{$^{4}$School of Gifted Young, University of Science and Technology of China, Hefei, Anhui 230026, China}
\begin{abstract}
Stochastic gravitational wave backgrounds~(SGWBs) receive increasing attention and provide a new possibility to directly probe the early Universe. In the preheating process at the end of inflation, parametric resonance can generate large energy density perturbations and efficiently produce gravitational waves~(GWs) which carry unique information about inflation. Since the peak frequency of such GWs is approximately proportional to the inflationary energy scale, $\Lambda_{\mathrm{inf}}$, GWs from preheating are expected to be observed by interferometer GW detectors in low-scale inflationary models. We investigate a class of preheating models where the effective potential has a quadratic minimum, and the dependence of the amplitude of such GWs on $\Lambda_{\mathrm{inf}}$, then find that the present energy spectrum of these GWs does not depend on $\Lambda_{\mathrm{inf}}$ only in the case of $\Lambda_{\mathrm{inf}}$ is above a critical value $\Lambda_{c}$, a parameter depending on the resonance strength. We numerically obtain $\Lambda_{c}$ in terms of the model parameters in linear approximation and then conduct lattice simulations to verify this result. For $\Lambda_{\mathrm{inf}}\lesssim\Lambda_{c}$, the amplitude of GWs quickly decreases with $\Lambda_{\mathrm{inf}}$ and becomes challenging to observe. In turn, observing such GWs in interferometer detectors also helps to determine $\Lambda_{\mathrm{inf}}$ and the resonance strength during the preheating.
\end{abstract}

\maketitle

\section{Introduction}
\label{sec:int}
Inflation is a successful model of the very early Universe which naturally solves the Horizon problem, the flatness problem, and the magnetic monopole problem simultaneously~\cite{Guth:1980zm, Linde:1983gd}. Primordial curvature perturbations from quantum fluctuations during inflation also successfully explain the CMB temperature fluctuations and seed the initial value of the large scale structure~\cite{Aghanim:2018eyx}. The existence of primordial GWs generated from quantum fluctuations of tensor modes is an important prediction of inflation, and is expected to be detected from B-mode polarization of CMB anisotropies~\cite{Seljak:1996gy}. Since the amplitude of primordial tensor perturbations depends on $\Lambda_{\mathrm{inf}}$ (the energy scale of inflation), 
the upper limits of tensor perturbations help to distinguish the inflationary models~\cite{Liddle:1993ch, Guo:2010mm}. The current constraint from CMB data on the tensor-to-scalar ratio is $r<0.09$ at $95\%$ level, excluding the models with quartic and cubic potentials~\cite{Akrami:2018odb}. 

Apart from primordial GWs, another prediction of GWs comes from the preheating process at the end of inflation~\cite{Khlebnikov:1997di}.
To set the initial conditions of the hot Big-Bang Universe, the vacuum energy transfers into radiation and reheats the universe after inflation, which is referred to as reheating~\cite{Albrecht:1982mp, Shtanov:1994ce}. Many inflationary models predict the existence of preheating process at the beginning of reheating, where  the perturbations of the inflaton are amplified exponentially by parametric resonance, generating large energy density perturbations inside the Hubble horizon~\cite{Shtanov:1994ce,Kofman:1997yn, Zhu:2018smk}. During the preheating, the equation of state of the Universe, $\omega$, could deviate from $1/3$~(in radiation-dominated Universe) or $0$~(in matter-dominated Universe), depending on the form of the effective potential. Therefore, the dynamics during the preheating affect the model prediction of the $e$-folding numbers of inflation, the amplitude of the power spectrum of scalar perturbations and the scalar spectral index~\cite{Lozanov:2016hid}.
In general, preheating is expected to happen at an energy scale much higher than that  colliders could reach~\cite{Starobinsky:1980te,Mukhanov:1981xt,Linde:1983gd}. GWs generated during the preheating then provide us a new opportunity to explore the end of inflation and the history of the early Universe. 

The detailed dynamics of preheating is investigated by analytical and numerical methods in various models~\cite{Greene:1997fu,Felder:2000hj,Sa:2007pc,Zhou:2013tsa,Lozanov:2019ylm,Sang:2019ndv,Wang:2021nuj,Hiramatsu:2020obh,Figueroa:2020rrl,Antusch:2017flz,Kost:2021rbi}. The case the inflaton $\phi$ is coupled to a scalar field $\chi$ by the term $\frac{1}{2}g^{2}\phi^{2}\chi^{2}$ is studied thoroughly in Ref.~\cite{Kofman:1997yn}. Ref.~\cite{Dufaux:2008dn} considers GWs from tachyonic preheating after hybrid inflation. Refs.~\cite{Adshead:2018doq,Cuissa:2018oiw,Adshead:2019lbr} consider the case the inflaton is coupled to gauge fields through the axion-like coupling. Refs.~\cite{Liu:2017hua, Liu:2018rrt} consider the amplification of perturbations and the formation oscillons in cuspy models. The parametric resonance induced by non-minimal coupling is investigated in Refs.~\cite{Fu:2017ero,GarciaBellido:2008ab, Jin:2020tmm}. Since the comoving Hubble horizon is very small at the end of inflation, the peak frequency of such GWs is in general much higher than the sensitive frequency of the current laser interferometers~\cite{Easther:2006vd}. For example, if $\Lambda_{\mathrm{inf}}$ is close to the grand unified theory scale, $10^{16}\mathrm{GeV}$, the peak frequency is above $10^{8}\mathrm{Hz}$. For the recent progress of detecting high-frequency GWs, see Refs.~\cite{Domcke:2020yzq,Aggarwal:2020olq}.
In particular, in hybrid inflation~\cite{Linde:1993cn} and curvaton models~\cite{Lyth:2001nq}, the inflationary potential is free from the CMB constraints so that $\Lambda_{\mathrm{inf}}$ could be much smaller. In particular, taking into account the trans-Planckian censorship conjecture~\cite{Bedroya:2019snp}, $\Lambda_{\mathrm{inf}}$ should be smaller than $10^{9}\mathrm{GeV}$, and then the peak frequency lies in the sensitivity bands of interferometer detectors such as aLIGO~\cite{TheLIGOScientific:2014jea}, DECIGO~\cite{Kawamura:2011zz}, LISA~\cite{Audley:2017drz}, Taiji~\cite{Guo:2018npi} and so on.

In Ref.~\cite{Easther:2006vd}, the authors simulate GWs from preheating and find the peak value of the GW energy spectrum does not depend on $\Lambda_{\mathrm{inf}}$. In this work, we revisit  this issue in detail and find that this conjecture is valid only in the case that  the parametric resonance is strong enough. We consider a broad class of models where the effective potentials have a quadratic minimum so that the resonance strength decreases with time. Since the initial value of perturbations decreases as $\Lambda_{\mathrm{inf}}^{2}$, perturbations of inflaton might remain much smaller than the background value throughout preheating in low-scale inflationary models. In this case, the energy density of perturbations is too small to generate considerable GWs. We study how the amplitude of GWs from the preheating depends on $\Lambda_{\mathrm{inf}}$ using both analytical and numerical methods in this work.

The paper is organized as follows. In Sec.~\ref{sec:mo}, we introduce the inflationary model investigated in this work. In Sec.~\ref{sec:li}, we analyze the dynamics of preheating in the linear approximation. In Sec.~\ref{sec:nu}, we show the results of the present energy spectrum of GWs from lattice simulation and compare them with the results in linear analysis. In Sec.~\ref{sec:con}, we summarize the results.
We set $c=8\pi G = 1$ throughout the paper.

\section{Models}
In this section, we briefly introduce the model investigated in this work.

Consider the model realized by the following action, 
\label{sec:mo}
\begin{equation}
	S=\int d^{4} x \sqrt{-g} \left(  -\frac{1}{2} R +\dfrac{1}{2}\partial_{\mu}\phi\partial^{\mu}\phi   +V(\phi)\right)\,,
\end{equation}
where $R$ is the Ricci scalar.
We investigate a class of models where the effective potential has a quadratic minimum. In this paper we take the $\alpha$-attractor E-model~\cite{Kallosh:2013hoa,Galante:2014ifa,Kallosh:2015lwa} as an example, where the effective potential reads
\begin{equation}
V(\phi)=V_{0}\left[1-\exp \left(-\frac{\phi}{M}\right)\right]^{2}\,.
\label{eq:Emodel}
\end{equation}
For single field inflaton, $V_{0}$ and $M$ are not free parameters but determined by the CMB observations. In this work, we consider two cases

1. Hybrid inflation models. In this case, the inflaton generates primordial perturbations and $\phi$ is the waterfall field relevant to preheating.

2. Curvaton models. In this case, $\phi$ is the inflaton, and the curvaton is another scalar field that generates primordial perturbations.

In the two cases, another scalar field is responsible to generate primordial perturbations and the potential of $\phi$ is free from CMB constraints. Thus, $V_{0}$ and $M$ are free parameters. 
Note that the condition $V(\phi)\sim\phi^{2}$ at $\phi\rightarrow 0$ is a common feature naturally realized in many models except potential~\eqref{eq:Emodel}, such as $\alpha$-attractor T-models and Monodromy models, also a coupling between $\phi$ and another scalar is allowed. Assuming the energy density during inflation is approximately a constant, then $\Lambda_{\mathrm{inf}}=V^{\frac{1}{4}}_{0}$. In the limit $\phi\rightarrow 0$, potential (\ref{eq:Emodel}) tends to be quadratic, $V(\phi)\simeq \frac{1}{2}m^{2}\phi^{2}$, where the effective mass $m=\sqrt{V_{0}}/M$.
\begin{figure}[t]
	\includegraphics[width=3in]{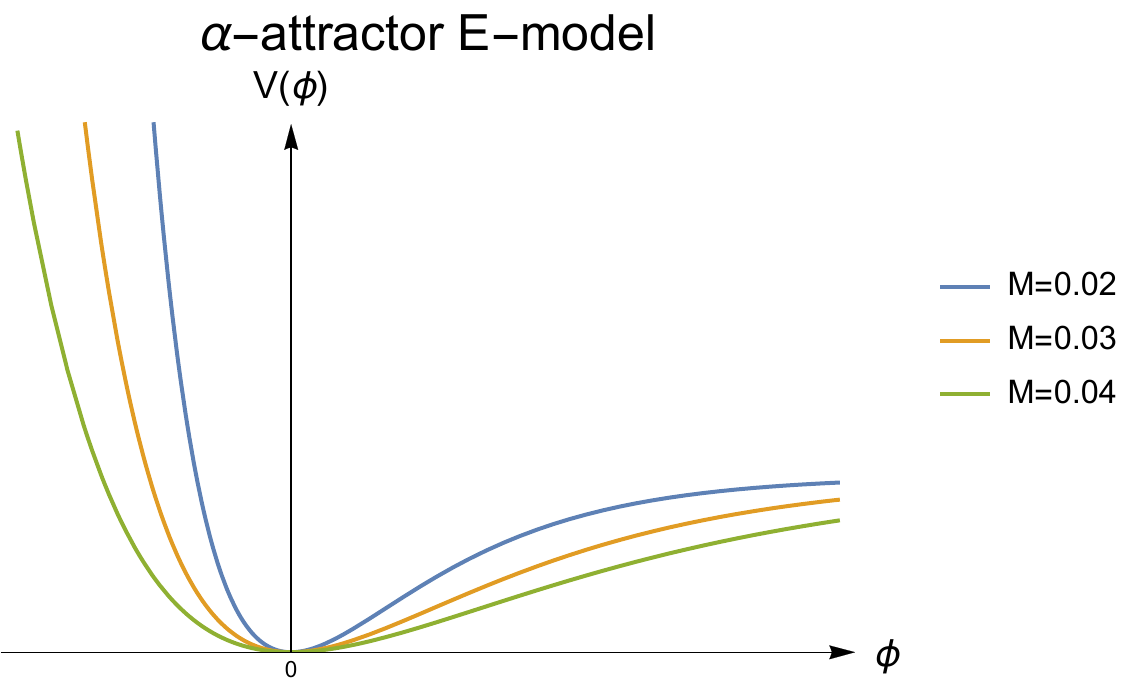}
	\caption{The effective potential of $\alpha$-attractor E-models.
	}
	\label{fig:pot}
\end{figure}


In a Friedmann-Lemaitre-Robertson-Walker Universe, the Friedman equation and the equation of motion~(EOM) of $\phi$ are
\begin{equation}
	\begin{split}
& H^2 = \frac{1}{3} \left\langle  \frac{1}{2}\dot{\phi}^2 + \frac{1}{2a^2}(\nabla\phi)^2 +V(\phi) \right\rangle \,, \\
& \ddot{\phi}-\frac{1}{a^{2}}\nabla^{2}\phi+3H\dot{\phi}+\frac{dV}{d\phi}=0\,,
\end{split}
\label{eq:motion}
\end{equation}
where $H$ is the Hubble parameter, $\langle...\rangle$ denotes a spatial average over the volume,
overdots denote derivatives with respect to the cosmic time $t$,
 and $\nabla$ is the spatial gradient. The quantity inside the angle bracket presents the energy density of $\phi$.
The EOM of tensor perturbations, $h_{ij}$, is derived from the linearized Einstein equation
\begin{equation}
	\label{eq:EoMhij}
	\ddot{h}_{ij}+3H\dot{h}_{ij}-\frac{1}{a^{2}}\nabla^{2}h_{ij}=\frac{2}{a^{2}} T_{ij}^{\mathrm{TT}}\,,
\end{equation}
where $T^\mathrm{TT}_{ij}$ is the transverse-traceless~(TT) component of the energy-momentum tensor $T_{ij}$, which takes the form
\begin{equation}
	\label{eq:Tij}
	T_{ij}=\partial_{i}\phi\partial_{j}\phi-\frac{1}{3}\delta_{ij}\partial_{k}\phi\partial^{k}\phi\,.
\end{equation}
In general, the resonance wavelength is more than two orders of magnitude smaller than the Hubble horizon scale, so scalar metric perturbations are negligible both in the EOMs of $\phi$ and $h_{ij}$.

The energy density of GWs is given by
\begin{equation}
	\label{eq:edgw}
	\rho_{\mathrm{GW}}=\frac{1}{4}\langle\dot{h}_{ij}\dot{h}^{ij}\rangle\,,
\end{equation}
and the dimensionless energy spectrum of GWs is defined by
\begin{equation}
\label{eq:es}
\Omega_\mathrm{GW}\equiv \frac{1}{\rho_c}\frac{d\rho_\mathrm{GW}}{d\ln k},
\end{equation}
where $\rho_c$ is the critical density of the Universe.

\section{Linear approximation}
\label{sec:li}
In this section, we investigate the dynamics of perturbations in the linear approximation and then give the results of  $\Lambda_{c}$ in terms of $M$ from the linear analysis.

After inflation, $\phi$ begins to oscillate around the minimum of its potential. At the beginning of preheating, $\phi$ is almost homogeneous with some small perturbations on it caused by quantum fluctuations. Thus, we split $\phi(\mathbf{x})$ as small fluctuations around a homogeneous field, $\phi(t,\mathbf{x})=\bar{\phi}(t)+\delta \phi(t,\mathbf{x})$. The EOMs of $\bar{\phi}$ and $\delta\phi$ are
\begin{equation}
	\label{eq:EOMbar}
	\ddot{\bar{\phi}}+3H\dot{\bar{\phi}}+\frac{dV}{d\bar{\phi}}=0\,,
\end{equation}
\begin{equation}
	\label{eq:EOMdelta}
	\ddot{\delta\phi_{k}}+\frac{k^{2}}{a^{2}}\delta\phi_{k}+3H\dot{\delta\phi_{k}}+\frac{d^2V}{d\bar{\phi}^2}\delta\phi_{k}=0\,,
\end{equation}
where $\delta\phi_{k}$ is the fourier form of $\delta\phi(\mathbf{x})$.

\begin{figure*}
	\includegraphics[height=2.6in]{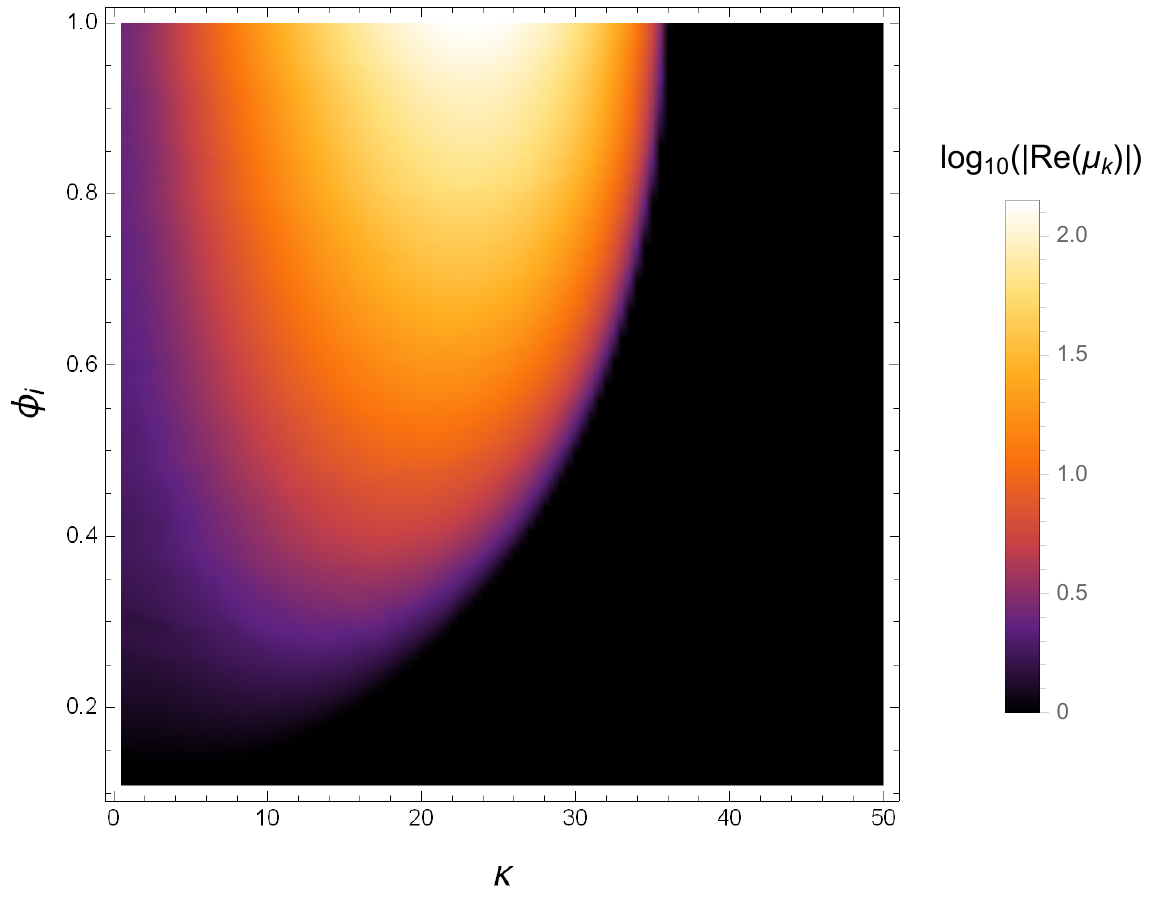}
	\includegraphics[height=2.6in]{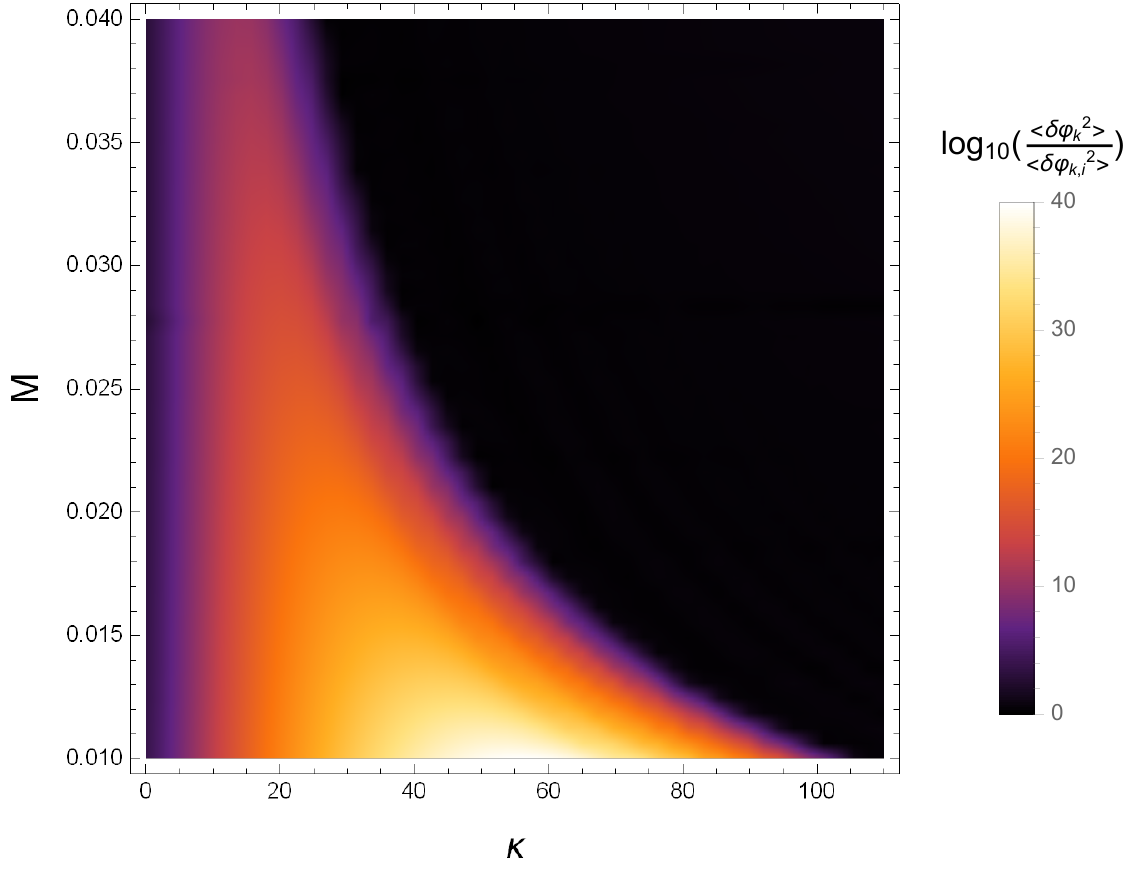}
	\caption{Left: The resonance strength $|\mathrm{Re}(\mu_{k})|$ in terms of $\phi_{i}$ and $\kappa$ in case of $M=0.02$. Right: $\langle\delta\varphi^{2}_{\mathbf{k}}\rangle/\langle\delta\varphi^{2}_{\mathbf{k},i}\rangle$ in terms of $M$ and $\kappa$ at the end of resonance.
	}
	\label{fig:listp}
\end{figure*}
Since the energy density of $\delta\phi$ is negligible at the beginning of preheating, the Hubble parameter is calculated from the energy density of $\bar{\phi}$. Thus, the EOM of $\bar{\phi}$ can be solved independently assuming field fluctuations have little effect on it.
In the linear analysis, each Fourier mode of fluctuations evolves independently
so that Eq.~\eqref{eq:EOMdelta} can be numerically solved as an ordinary differential equation.

As an illustration, in linear analysis we choose $\bar{\phi}_{i}=2M$ and $\dot{\bar{\phi}}_i=0$
as the initial conditions for the homogeneous field. The subscript $i$ denotes the initial value of the variables at the beginning of preheating throughout this paper.
The initial value of perturbations is obtained as the Bunch-Davies type
\begin{equation}
	\label{eq:BD}
	\delta\phi_{k,i}=\dfrac{1}{a\sqrt{2k}}e^{ik\eta}\,, 
\end{equation}
where $\eta$ is the conformal time. 


If we firstly neglect the expansion of the Universe, the oscillation amplitude of $\bar{\phi}$ is a constant so that Eq.~\eqref{eq:EOMdelta} is periodic in Minkowski space.
Then, according to the Floquet theory, the periodic equation~\eqref{eq:EOMdelta} has a general solution
\begin{equation}
	\label{eq:EOMFlo}
	\delta\phi_{k}=\mathcal{P}_{k+}(t)\exp(\mu_{k}t)+\mathcal{P}_{k-}(t)\exp(-\mu_{k}t)\,,
\end{equation}
where $\mu_{k}$ is the Floquet exponent and $\mathcal{P}_{k\pm}$ are periodic functions determined by initial conditions.

\begin{figure}
	\includegraphics[width=3in]{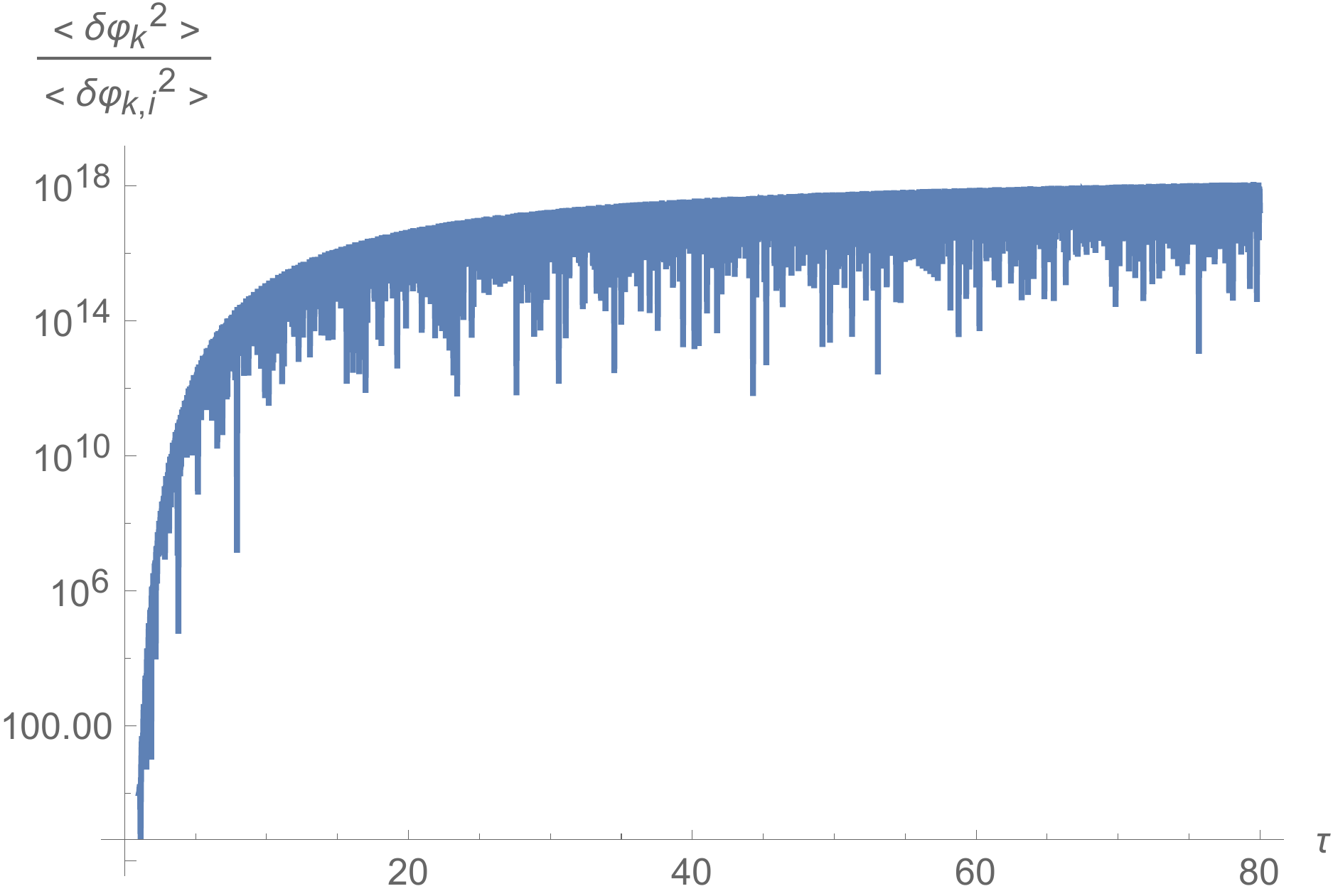}
	\caption{The amplification of $\langle\delta\varphi_{k}^{2}\rangle$, where we choose $\kappa=20$ and $M=0.02$.
	}
	\label{fig:grow}
\end{figure}

In that case that the real part of $\mu_{k}$ is nonzero, i.e., $\mathrm{Re}(\mu_{k}) \neq 0$,
perturbations are unstable and $\delta\phi_{k}$ grows exponentially.

We use the rescaled wavenumber $\kappa\equiv k/\sqrt{V_{0}}$ and rescaled time $\tau\equiv t\,\sqrt{V_{0}}$ so that the resonant modes satisfy $\kappa\sim\mathcal{O}(1)-\mathcal{O}(10^{3})$ and the preheating process sustains for about $\tau\sim \mathcal{O}(10^{2})$.
With the rescaled parameters we absorb $V_{0}$ in Eq.~\eqref{eq:EOMdelta} and the EOM reads
\begin{equation}
	\label{eq:1}
	\frac{d^{2}}{d\tau^{2}}\delta\phi_{k}+3\frac{da/d\tau}{a}\frac{d}{d\tau}\delta\phi_{k}+\left(\frac{d^{2}(V/V_{0})}{d\bar{\phi}^{2}}+\frac{\kappa^{2}}{a^{2}}\right)\delta\phi_{k}=0\,,
\end{equation}
which implies that the resonance bands are independent of $V_{0}$.
The left panel of Fig.~\ref{fig:listp} shows the dependence of $|\mathrm{Re}(\mu_{k})|$ on $\kappa$ and $\phi_{i}$, one can find that $|\mathrm{Re}(\mu_{k})|$ is close to zero for $\phi_{i}\ll M$. This is because the potential is proportional to $\phi^{2}$ at its bottom, then Eq.~\eqref{eq:EOMdelta} implies $\delta\phi$ could be treated as a free massive particle and the amplification of perturbations does not occur. As a result, the resonance strength decreases with the oscillation amplitude of $\bar{\phi}$ in the expanding Universe. For smaller $M$, the resonance is stronger because the effective mass $m$ becomes larger and $\bar{\phi}$ oscillates more times in unit time. One can find the resonance band is wider and $|\mathrm{Re}(\mu_{k})|$ is larger for smaller $M$ in the left panel of Fig.~\ref{fig:listp}. 

When the oscillation amplitude of $\bar{\phi}$ becomes much smaller than $M$, Eqs.~\eqref{eq:EOMbar} and~\eqref{eq:EOMdelta} suggest that both $\bar{\phi}$ and $\delta\phi$ behave as massive particles with effective mass $m$, and their amplitude decreases as $a^{-3/2}$. Then, we use $\bar{\varphi}\equiv a^{3/2}\bar{\phi}$ and $\delta\varphi\equiv a^{3/2}\delta\phi$ instead to include the effect of the expansion of the Universe.

In Fig.~\ref{fig:grow} we show the evolution of $\langle\delta\varphi^{2}_{\mathbf{k}}\rangle$, which implies $\langle\delta\varphi^{2}_{\mathbf{k}}\rangle$ increases for about $10^{18}$ and finally tends to be a constant. The right panel of Fig.~\ref{fig:listp} shows $\langle\delta\varphi^{2}_{\mathbf{k}}\rangle/\langle\delta\varphi^{2}_{\mathbf{k},i}\rangle$ in terms of $M$ and $\kappa$ at the end of resonance.

To quantify the amplitude of perturbations, we apply the varience of $\delta\varphi$, $\langle\delta\varphi^{2}\rangle$, which is defined as
\begin{equation}
	\label{eq:phisq}
	\langle\delta\varphi^{2}\rangle\equiv\dfrac{1}{\mathcal{V}}\int d^{3}x|\delta\varphi(\mathbf{x})|^{2}\approx\int^{k_{max}}_{k_{min}}\dfrac{d^{3}k}{(2\pi)^{3}}|\delta\varphi_{\mathbf{k}}|^{2}\,,
\end{equation}
where $\mathcal{V}$ is the volume of the integration region, and the range from $k_{min}$ to $k_{max}$ covers the main resonance bands\footnote{Here we focus on the effect of parameter resonance during preheating, where the divergent high frequency vacuum fluctuations do not contribute.}. The initial value of the varience of $\delta\varphi$, $\langle\delta\varphi^{2}_{i}\rangle$, can be obtained from Eq.~\eqref{eq:BD} and Eq.~\eqref{eq:phisq}
\begin{equation}
	\label{eq:sqi}
	\langle\delta\varphi_{i}^{2}\rangle\approx V_{0}\int_{\kappa_{min}}^{\kappa_{max}}\frac{\kappa d\kappa}{(2\pi)^{2}}\,,
\end{equation}
where we have set $a=1$ initially. Since the resonance band of $\kappa$ does not depend on $V_{0}$ as shown in Eq.~\ref{eq:1}, we safely apply the same $\kappa_{min}$ and $\kappa_{max}$ for different $\Lambda_{\mathrm{inf}}$, and so that the integration $\int_{\kappa_{min}}^{\kappa_{max}}\frac{\kappa d\kappa}{(2\pi)^{2}}$ is a constant. Then, Eq.~\eqref{eq:sqi} implies that $\langle\delta\varphi_{i}^{2}\rangle$ is proportional to $V_{0}$, i.e., the initial value of inflaton perturbations becomes smaller for lower inflationary scale, and the resonance need to be more violent to amplify perturbations to the background value.

\begin{figure*}[t]
	\includegraphics[width=3in]{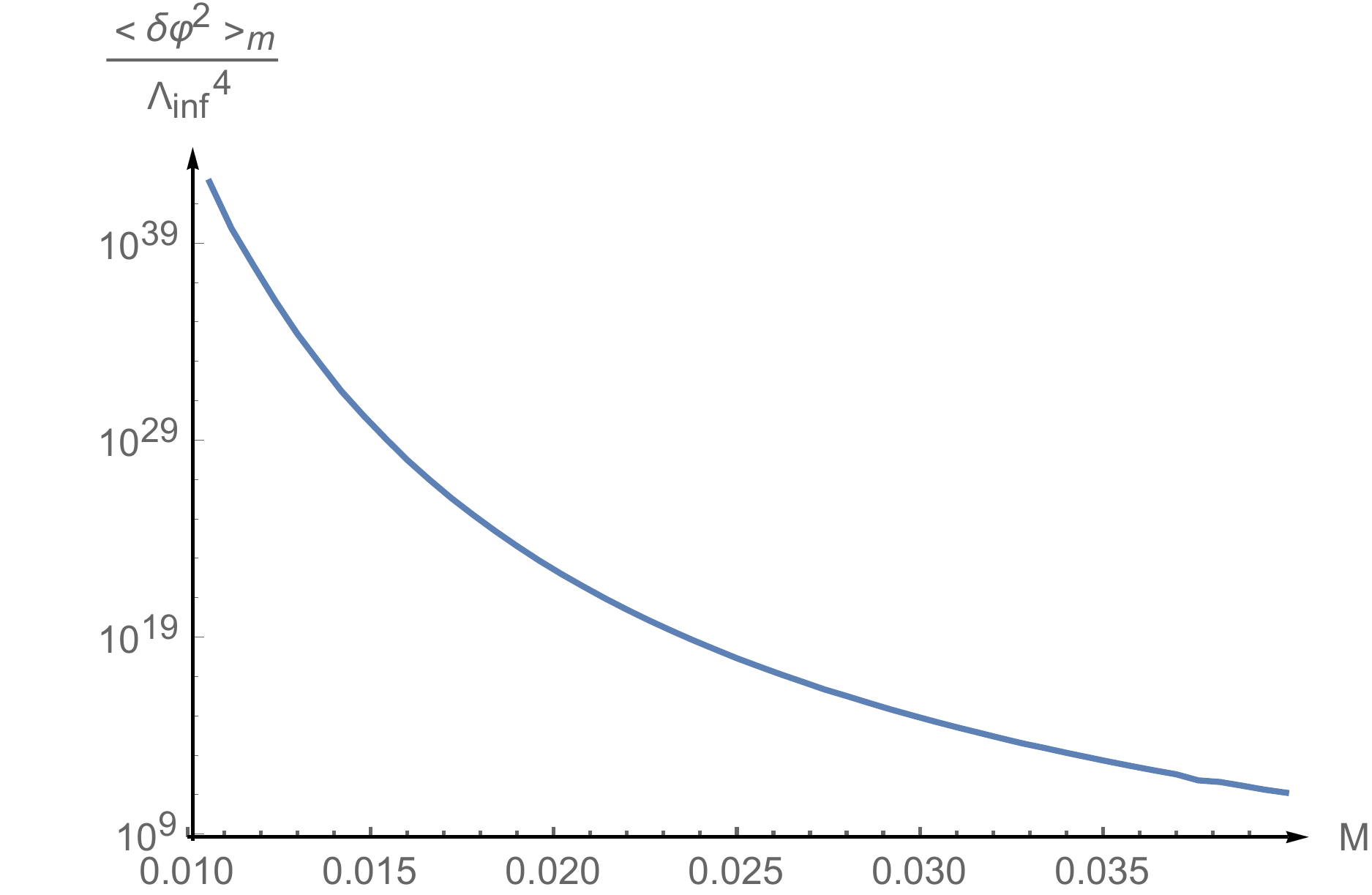}
	\includegraphics[width=3in]{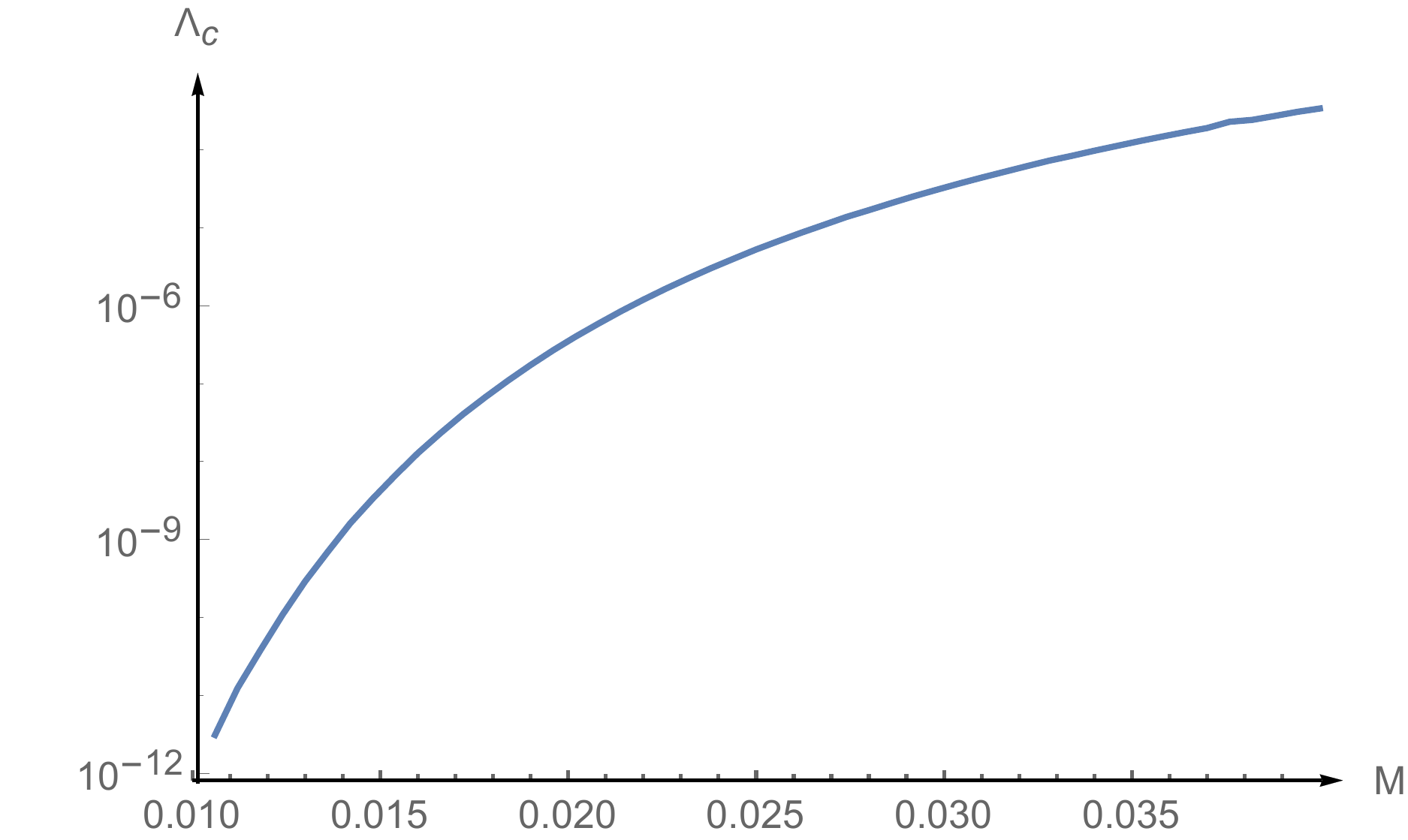}
	\caption{Left: $\langle\delta\phi^{2}\rangle_{m}/\Lambda_{\mathrm{inf}}^{4}$ in terms of $M$. Right: $\Lambda_{c}$ in terms of $M$ from linear analysis.
	}
	\label{fig:lin}
\end{figure*}

Since the resonance is only efficient in a limited time, the amplification of $\langle\delta\varphi^{2}\rangle$ during preheating has an upper bound.
In other words, if we put aside the nonlinear effects and solve Eq.~\eqref{eq:EOMdelta} under the linear approximation, there is a maximum value $\langle\delta\varphi^{2}\rangle$ which could finally reach at the end of resonance, denoted as $\langle\delta\varphi^{2}\rangle_{m}$. But the linear approximation is violated when $\delta\phi$ is comparable to $\bar{\varphi}$. At that time, the oscillation amplitude of $\bar{\phi}$ quickly decreases, Eq.~\eqref{eq:EOMdelta} is violated and we need to conduct lattice simulations to capture the nonlinear dynamics of preheating. Depending on the value of $\delta\phi_{i}$, the system will become nonlinear at different times or remains in the linear stage if $\delta\phi_{i}$ is too small. We assume the system becomes nonlinear when $\langle\delta\phi^{2}\rangle$ exceeds $\bar{\phi}^{2}$.
The case is then referred to as sufficient/insufficient preheating when  the evolution of the system  becomes nonlinear/remains linear at the end of resonance. 
The left panel of Fig.~\ref{fig:lin} shows that $\langle\delta\varphi^{2}\rangle_{m}/\Lambda_{\mathrm{inf}}^{4}$ increases inversely with $M$, which agrees with the results shown in Fig.~\ref{fig:listp} that $|\mathrm{Re}(\mu_{k})|$ is larger for smaller $M$. Then, the threshold $\Lambda_{c}$ can be obtained by considering the case $\delta\phi$ just reaches  the nonliear condition $\langle\delta\varphi^{2}\rangle_{m}=\varphi_{i}^{2}$ at the end of resonance, as shown in the right panel of Fig.~\ref{fig:lin}.

As for the amplitude of GWs, it is natural to expect that in the sufficient preheating case the energy density perturbations are amplified to the maximum value by nonlinear evolution, so that the amplitude of GWs remains the same. 
In the insufficient preheating case, most of the energy remains storing in the homogeneous part of the scalar field at the end of resonance. Since GWs are generated by the TT part of the energy density, without sufficient amplification of perturbations, the GW production is expected to be suppressed.

Then, we focus on the analysis of $\Omega_{\mathrm{GW}}$. In the rescaled coordinate $\tau\equiv t\sqrt{V_{0}}$ and $x^{I}\equiv x^{i}\sqrt{V_{0}}$, The EOMs of $\phi$ and $h_{ij}$ read
\begin{equation}
	\label{eq:eomphi2}
	\partial_{\tau}^{2}\phi-\frac{1}{a^{2}}\partial_{I}\partial^{I}\phi+3\frac{da/d\tau}{a}\partial_{\tau}\phi+\frac{d(V/V_{0})}{d\phi}=0\,,
\end{equation}
\begin{equation}
	\label{eq:eomhij2}
	\partial_{\tau}^{2}h_{ij}+3\frac{da/d\tau}{a}\partial_{\tau}h_{ij}-\frac{1}{a^{2}}\partial_{I}\partial^{I}h_{ij}=\frac{2}{a^{2}}(\partial_{I}\phi\partial_{J}\phi)^{\mathrm{TT}}\,,
\end{equation}
In the form of Eqs~\eqref{eq:eomphi2} and \eqref{eq:eomhij2}, one can find that all the EOMs do not depends on $V_{0}$. The dependence of $\Omega_\mathrm{GW}$ on $\Lambda_{\mathrm{inf}}$ is only caused by the initial value of quantum fluctuations, $\langle\delta\phi_{i}^{2}\rangle\propto V_{0}$, as shown in Eq.~\eqref{eq:sqi}. Eq.~\eqref{eq:eomhij2} implies that $h_{ij}$ is proportional to the source term, $\frac{2}{a^{2}}(\partial_{I}\phi\partial_{J}\phi)^{\mathrm{TT}}$, which is quadratically depend on $\delta\phi$. We further assume the source term is simply proportional to $\langle\delta\phi^{2}\rangle$.
Then, the energy fraction of GWs produced during preheating, $\overline{\Omega_\mathrm{GW}}\equiv \rho_{\mathrm{GW}}/\rho_{c}$, is obtained by
\begin{equation}
	\label{eq:es2}
	\overline{\Omega_\mathrm{GW}}= \frac{1}{4}\frac{\left\langle\partial_{ \tau}h_{ij}\partial_\tau h^{ij}\right\rangle}{\rho_{c}}\propto\langle\delta\phi_{m}^{2}\rangle^{2}\,,
\end{equation}
where $\rho_{c}$ has been estimated as $V_{0}$ at the end of inflation.
In sufficient preheating case, $\sqrt{\langle\delta\phi_{m}^{2}\rangle}$ could reach the same order of $\bar\phi$, so that $\overline{\Omega_{\mathrm{GW}}}$ is a constant and independent of $\Lambda_{\mathrm{inf}}$.
In insufficient preheating case, since the amplification rate of $\langle\delta\phi^{2}\rangle$ is a constant, i.e., $\langle\delta\phi^{2}\rangle_{m}\propto\langle\delta\phi_{i}^{2}\rangle$, $\overline{\Omega_{\mathrm{GW}}}$ is further proportional to $\langle\delta\phi_{i}^{2}\rangle^{2}$.
Using the relation $\langle\delta\phi_{i}^{2}\rangle\propto \Lambda_{\mathrm{inf}}^{4}$, we can obtain the dependence of $\overline{\Omega_{\mathrm{GW}}}$ on $\Lambda_{\mathrm{inf}}$ as
\begin{equation}
	\label{eq:Omegalambdainf}
	\overline{\Omega_{\mathrm{GW}}}=\left\{
	\begin{aligned}
		&\Omega_{\mathrm{max}},&\text{for}\;\Lambda_{\mathrm{inf}}\geq\Lambda_{c},\\
		&\Omega_{\mathrm{ins}}(\Lambda_{\mathrm{inf}}/\Lambda_{c})^{8}, &\Lambda_{\mathrm{inf}}\ll\Lambda_{c},\\
	\end{aligned}
	\right.
\end{equation}
where $\Omega_{\mathrm{max}}$ and $\Omega_{\mathrm{ins}}$ are the maximum values that $\Omega_{\mathrm{GW}}$ could reach in the sufficient preheating case and the insufficient preheating case, respectively. Note that the derivation of Eq.~(17) does not contain any features of the $\alpha$-attractor $E$-model. The result Eq.~(17) is applicable to the models where the resonance gradually quenches as the oscillation amplitude of $\bar{\phi}$ decreases with time so that the total amplification rate of $\langle\delta\varphi^{2}\rangle$ is a finite value. Intuitively speaking, GWs should be stronger in sufficient preheating case, i.e., $\Omega_{\mathrm{max}}\geq\Omega_{\mathrm{ins}}$. For $\Lambda_{\mathrm{inf}}>\Lambda_{c}$, $\Omega_{\mathrm{GW}}$ is independent of $\Lambda_{\mathrm{inf}}$. For $\Lambda_{\mathrm{inf}}<\Lambda_{c}$, $\Omega_{\mathrm{GW}}$ quickly decreases and becomes more difficult to be observed.  This analytical result will be  compared with the lattice simulation result in the next section. Note that in Eq.~\eqref{eq:Omegalambdainf} there is a huge difference between $\Omega_{\mathrm{max}}$ and $\Omega_{\mathrm{ins}}$, which implies $\Omega_{\mathrm{GW}}$ has a sudden change near $\Lambda_{\mathrm{inf}}=\Lambda_{c}$, which is also stressed in the next section.


\section{Lattice simulations}
\label{sec:nu}
In this section, we present the numerical methods used in lattice simulation, and the numerical results of the present energy spectrum of GWs, $\Omega_{\mathrm{GW},0}$, for different $M$ and $\Lambda_{\mathrm{inf}}$. Then, we show the dependence of $\Omega_{\mathrm{GW},0}$ on $\Lambda_{\mathrm{inf}}$ from lattice simulation, and compare with the results in the linear analysis.

When perturbations become comparable to the homogeneous part, nonlinear effects cannot be neglected and the evolution of $\phi$ has to be solved in lattice simulations.
To improve computational accuracy of the simulations, we apply the redefined variables as in LATTICEEASY~\cite{Felder:2000hq} and DEFROST~\cite{Frolov:2008hy}, which are given by 
\begin{equation}
	\label{eq:propara}
	\begin{split}
		\phi_{\mathrm{pr}} = a^{3/2}\phi/\phi_{i}\,,\quad
		dt_{\mathrm{pr}} = dt\sqrt{V_{0} }/M\,,\quad
		k_{\mathrm{pr}} = k\,M/\sqrt{V_{0} }\,.
	\end{split}
\end{equation}
We apply the finite-difference method to solve the EOMs of $\phi$ and $h_{ij}$ in configuration space, both the spacial derivatives and the time derivatives are realized using the fourth-order method with double precision. 
Also, we use a performance-portable parallel programming model, OPENACC, to accelerate the code with GPUs.

To reduce the computational cost of the simulation,
we have defined a new quantity $u_{ij}$, and the EOM of $u_{ij}$ reads
\begin{equation}
	\label{eq:EoMuij}
	\ddot{u}_{ij}+3H\dot{u}_{ij}-\frac{1}{a^{2}}\nabla^{2}u_{ij}=\frac{2}{a^{2}}T_{ij}\,.
\end{equation}
Then, $h_{ij}$ can be obtained by conduct the TT projection on $u_{ij}$
\begin{equation}
	h_{ij}(t,{\bf k})=\Lambda_{ij,lm}(\hat{\bf k})u_{lm}(t,{\bf k}),
\end{equation}
where $u_{lm}(t,{\bf k})$ is the Fourier form of $u_{ij}$ and the projection operator $\Lambda_{ij,lm}(\hat{\bf k})$ reads
\begin{equation}
	\Lambda_{ij,lm}(\hat{\bf k}) \equiv P_{il}(\hat{\bf k})P_{jm}(\hat{\bf k}) - \frac12 P_{ij}(\hat{\bf k})P_{lm}(\hat{\bf k}),
\end{equation}
with $P_{ij} \equiv \delta_{ij}-\hat{k}_i \hat{k}_j$.
Thus, one needs to conduct the $\mathrm{TT}$ projection on $u_{ij}$ only at the times exporting the value of $\Omega_{\mathrm{GW}}$, and avoids the calculation of the $\mathrm{TT}$ projection at each step.

Similar to the EOM of the $\phi$, we thus can evolve Eq.~\eqref{eq:EoMuij} in configuration space in the simulations.
In terms of $u_{ij}$, $\Omega_{\mathrm{GW}}$ then can be expressed as~\cite{Figueroa:2011ye}
\begin{equation}
\Omega_{\mathrm{GW}}
=\frac{k^3}{4L^3\rho_c}\int d \Omega\, \Lambda_{ij,lm}(\hat{\mathbf{k}}) \dot{u}_{ij}(t,\mathbf{k})\dot{u}_{lm}^{*} (t,\mathbf{k})\,.
\end{equation}

The expansion rate of the Universe is calculated self-consistently from spatially averaged energy density.
We perform three-dimensional lattice simulations with $256^{3}$ points in a box with periodic boundary conditions.
The size of the box $L$ and the number of grid points per edge $N$ are in principle chosen
according to the physical features of the model.
In our simulations, the box size is chosen to be close to the resonance wavelength,
which is smaller than the Hubble horizon size,
so that the interesting wavelengths such as the physical peaks in $\Omega_{\mathrm{GW}}$ are
located comfortably in between
the largest wavelength $\sqrt{3}L$~(diagonal line) and the smallest wavelength $L/N$.

\begin{figure}[t]
	\includegraphics[width=3in]{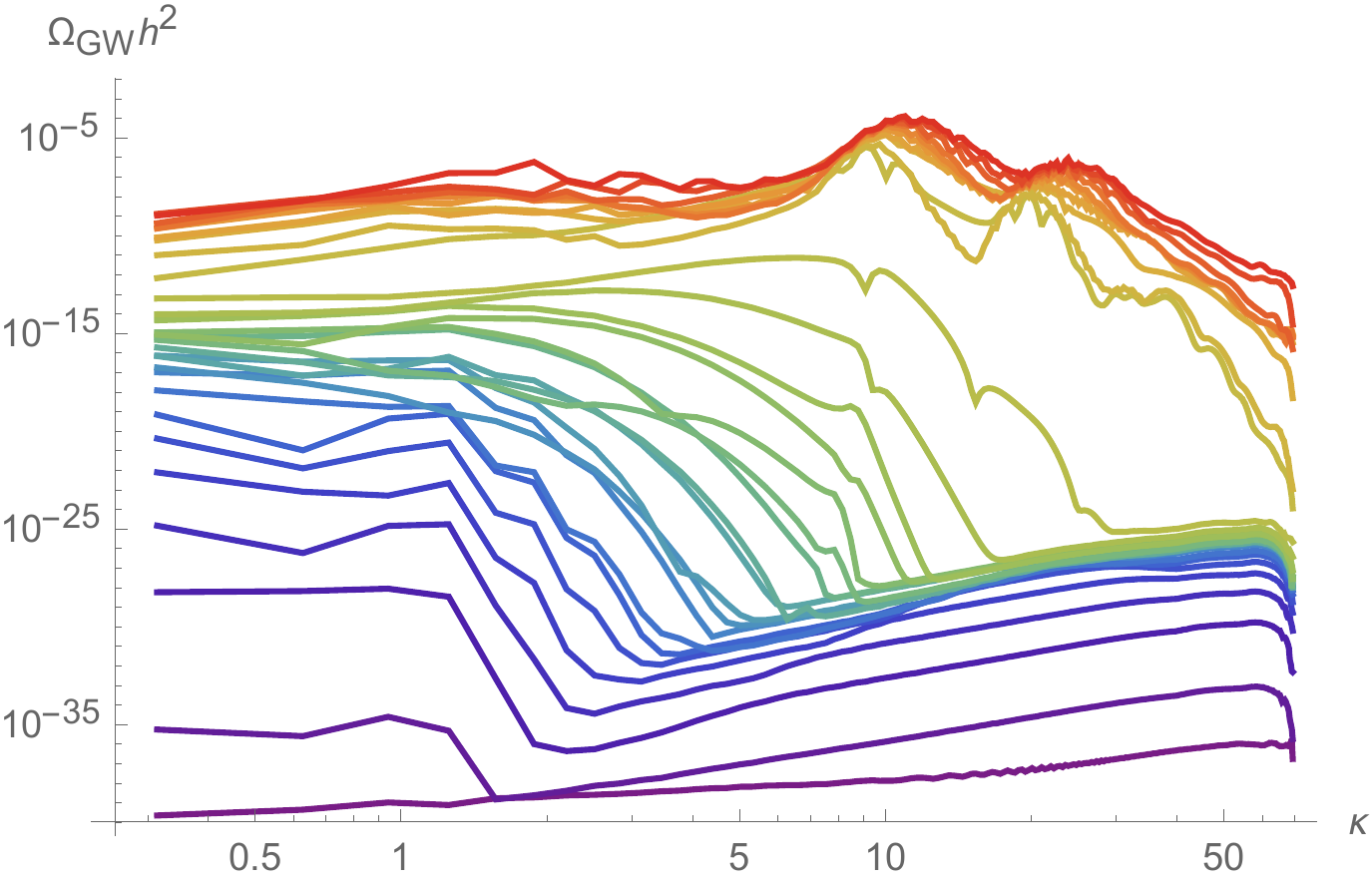}
	\caption{The evolution of $\Omega_{\mathrm{GW}}$ in case of $M=0.02$ and $\Lambda_{\mathrm{inf}}=3.8\times 10^{-7}$. The data is exported for every $\Delta\tau=100$ and the program ends at $\tau=3000$ when $\Omega_{\mathrm{GW}}$ stops increasing.
	}
	\label{fig:evo}
\end{figure}

The initial values of $\bar{\phi}$ and $\delta\phi$ are the same as in Sec.~\ref{sec:li}.
The evolution of $\Omega_{\mathrm{GW}}$ from lattice simulations is shown in Fig.~\ref{fig:evo}, where $\Omega_{\mathrm{GW}}$ is exported every $\Delta\tau=100$ and the simulation ends at $\tau=3000$ when $\Omega_{\mathrm{GW}}$ stops increasing. The evolution of $\Omega_{\mathrm{GW}}$ has experienced roughly three stages, as also mentioned in Ref.~\cite{Zhou:2013tsa}. Firstly, the increase of $\Omega_{\mathrm{GW}}$ appears only  in the low-$k$ region, corresponding to the resonant amplification of perturbations in the linear stage. Secondly, the nonlinear effects starts to dominate the evolution, 
and in the high-$k$ region $\Omega_{\mathrm{GW}}$ quickly increases. Thirdly, the nonlinear evolution is fully established, the compact objects, oscillons form~\cite{Amin:2010jq,Gleiser:2009ys,Muia:2019coe,Kou:2019bbc,Nazari:2020fmk} and $\Omega_{\mathrm{GW}}$ gradually stops increasing. The two peaks of $\Omega_{\mathrm{GW}}$ appear in the high-$k$ region as a consequence of oscillons.

To estimate $\Omega_{\mathrm{GW},0}$ and the corresponding frequency $f$, we assume the thermal equilibrium is established shortly after the end of the simulation.
Since the energy density of radiation evolves as $\rho_{r}\propto g^{-1/3}a^{-4}$,
$\Omega_{\mathrm{GW},0}$ and the frequency $f$ read
\begin{equation}
	\label{eq:GWnow}
	\Omega_{\mathrm{GW},0}
	=\Omega_{r,0}\left(\frac{g_{0}}{g_{*}}\right)^{1/3}\Omega_{\mathrm{GW}}\,,
\end{equation}

\begin{equation}
	\label{eq:fnow}
	f\simeq  \frac{k}{a_{e}\Lambda_{\mathrm{inf}}} \left(\frac{g_0}{g_*}\right)^{1/12} 4 \times 10^{10}\;\mathrm{Hz}\,,
\end{equation}
where $\Omega_{r,0}$ is the density fraction of radiation at present, $a_{e}$ is the scale factor at the end of the simulation,
and $g_{0}$ and $g_{e}$ are the effective numbers of ultrarelativistic degree of freedoms at present and at the end of the simulation, respectively.
\begin{figure*}
	\includegraphics[width=6in]{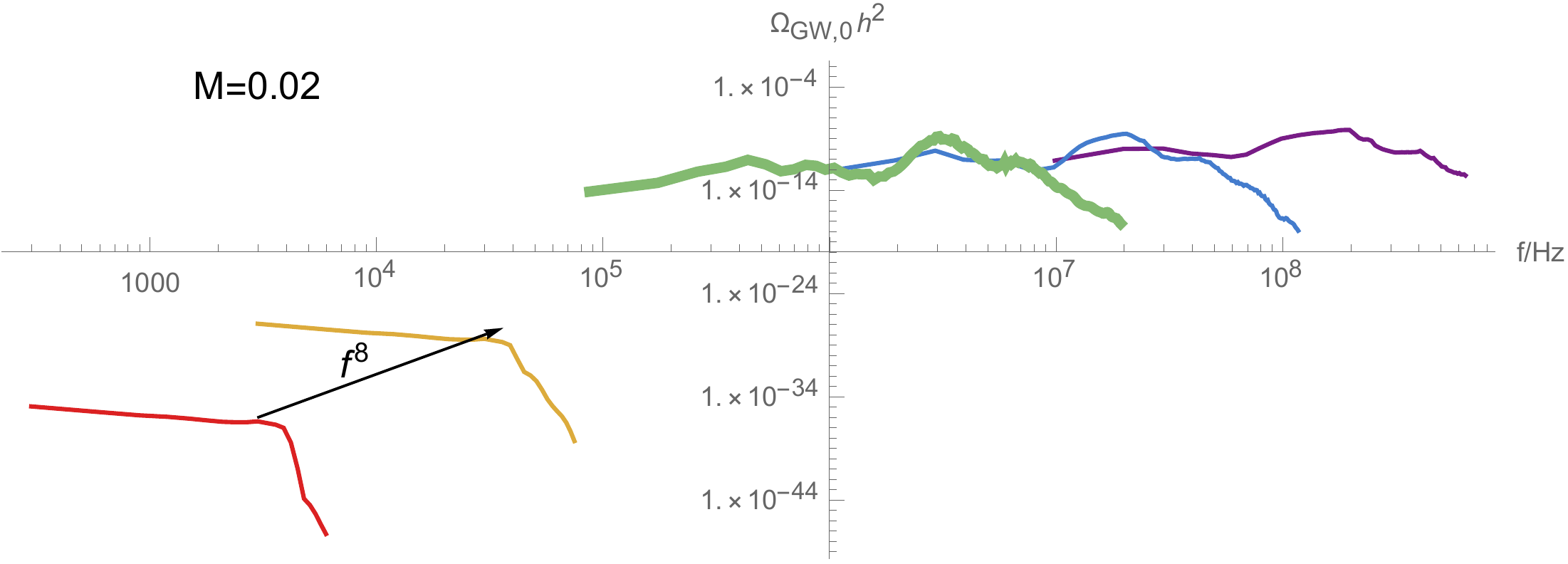}
	\raisebox{0.7\height}{\includegraphics[width=1in]{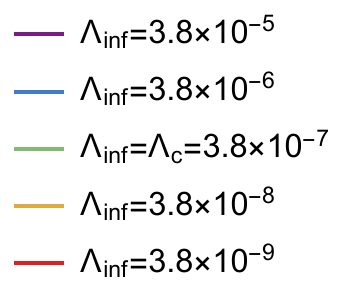}}
	\includegraphics[width=6in]{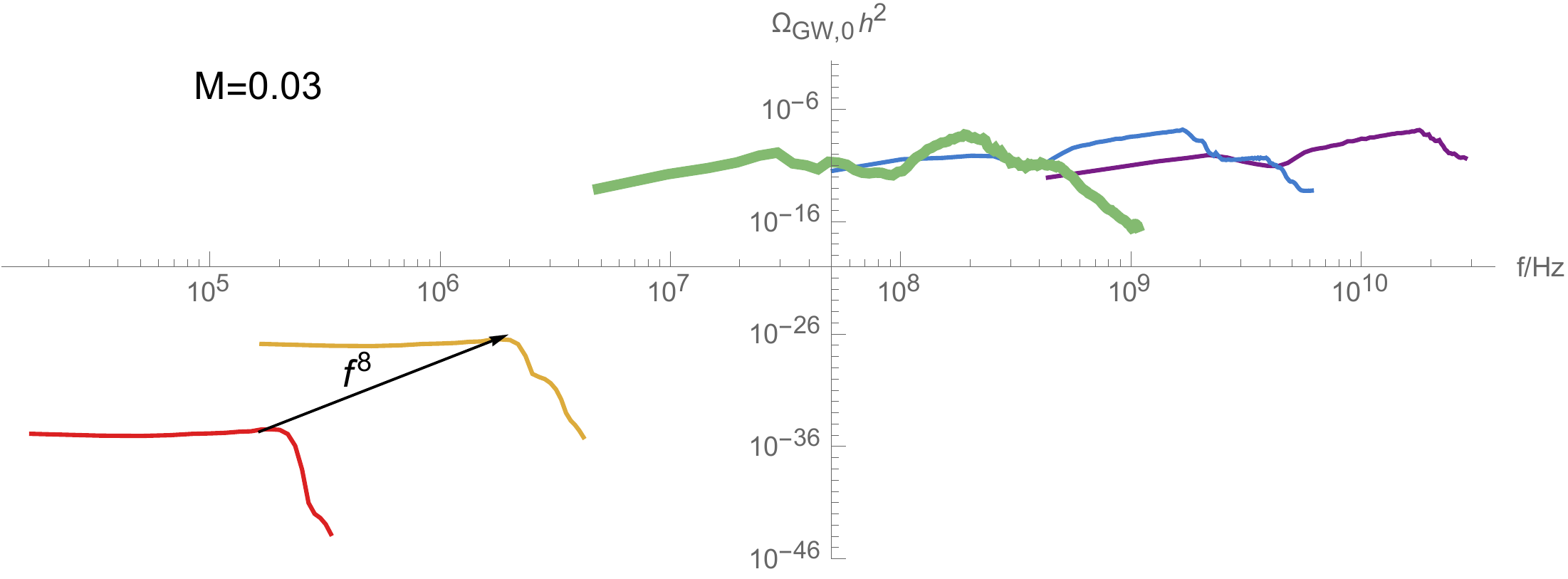}
	\raisebox{0.7\height}{\includegraphics[width=1in]{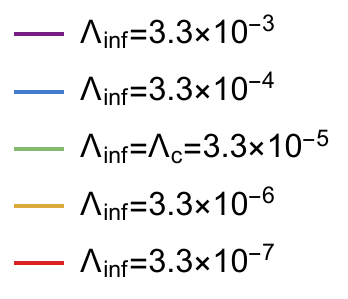}}
	\includegraphics[width=6in]{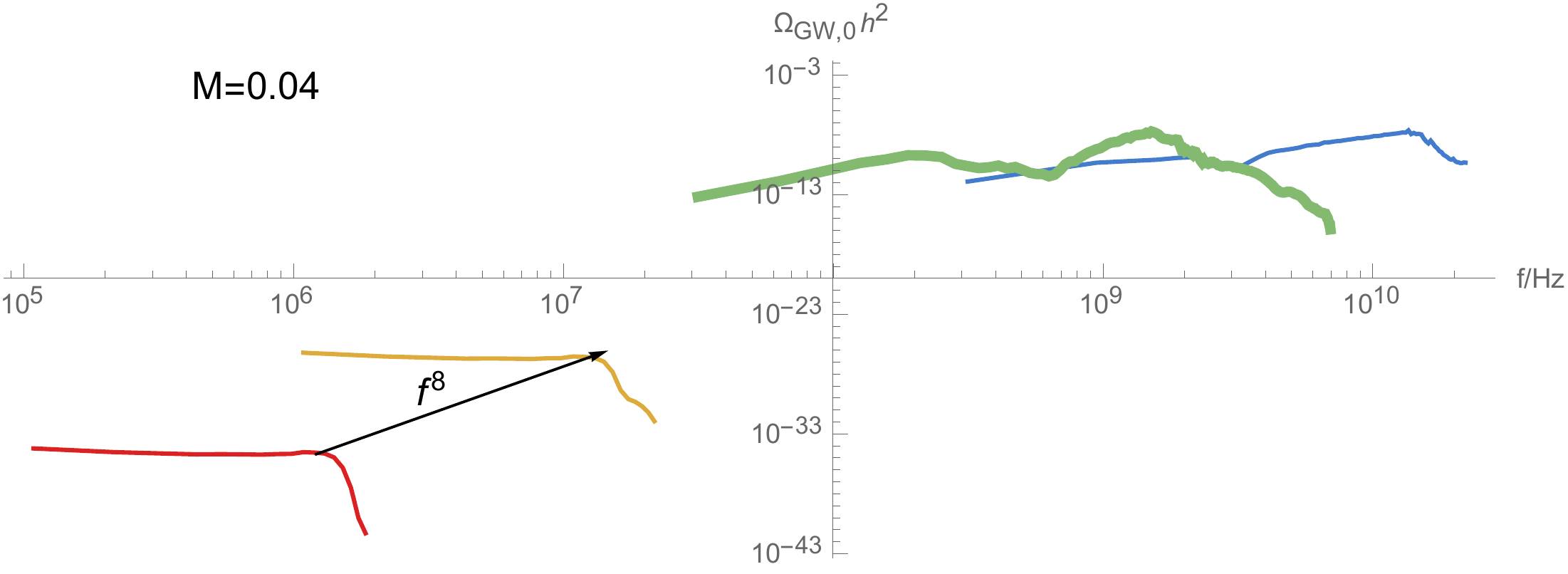}
	\raisebox{0.9\height}{\includegraphics[width=1in]{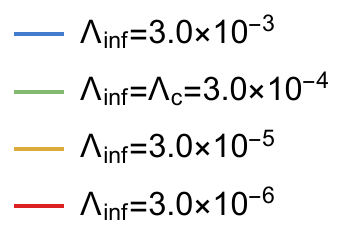}}
	\caption{The dependence of $\Omega_{\mathrm{GW},0}$ on $\Lambda_{\mathrm{inf}}$ for $M=0.02$, $M=0.03$ and $M=0.04$ case, where the thick green line in each panel denotes the result for $\Lambda_{\mathrm{inf}}=\Lambda_{c}$ predicted in the Sec.~\ref{sec:li}.
	}
	\label{fig:Mfre}
\end{figure*}

In Fig.~\ref{fig:Mfre} we show the dependence of $\Omega_{\mathrm{GW},0}$ on $\Lambda_{\mathrm{inf}}$ in the case of $M=0.02$, $M=0.03$, $M=0.04$, respectively. The values of $\Lambda_{c}$ are obtained from the linear analysis in Sec.~\ref{sec:li}. For $M=0.02$, $\Lambda_{c}=3.8\times 10^{-7}$, for $M=0.03$, $\Lambda_{c}=3.3\times 10^{-5}$, for $M=0.04$, $\Lambda_{c}=3.0\times 10^{-4}$, respectively. The green thick line in each panel of Fig.~\ref{fig:Mfre} shows the predicted $\Omega_{\mathrm{GW},0}$ for $\Lambda_{\mathrm{inf}}=\Lambda_{c}$.
Fig.~\ref{fig:Mfre} shows that the peak values of $\Omega_{\mathrm{GW},0}$ are almost constant for $\Lambda_{\mathrm{inf}}>\Lambda_{c}$, corresponding to the sufficient preheating case, and quickly decreases for $\Lambda_{\mathrm{inf}}<\Lambda_{c}$, corresponding to the insufficient preheating case. The peak frequency is roughly proportional to $\Lambda_{\mathrm{inf}}$, and $\Omega_{\mathrm{GW},0}$ is proportional to $f^{8}$ for $\Lambda_{\mathrm{inf}}<\Lambda_{c}$. Thus, the lattice results verify the result~\eqref{eq:Omegalambdainf} in the linear analysis.

\begin{figure*}
	\includegraphics[width=5in]{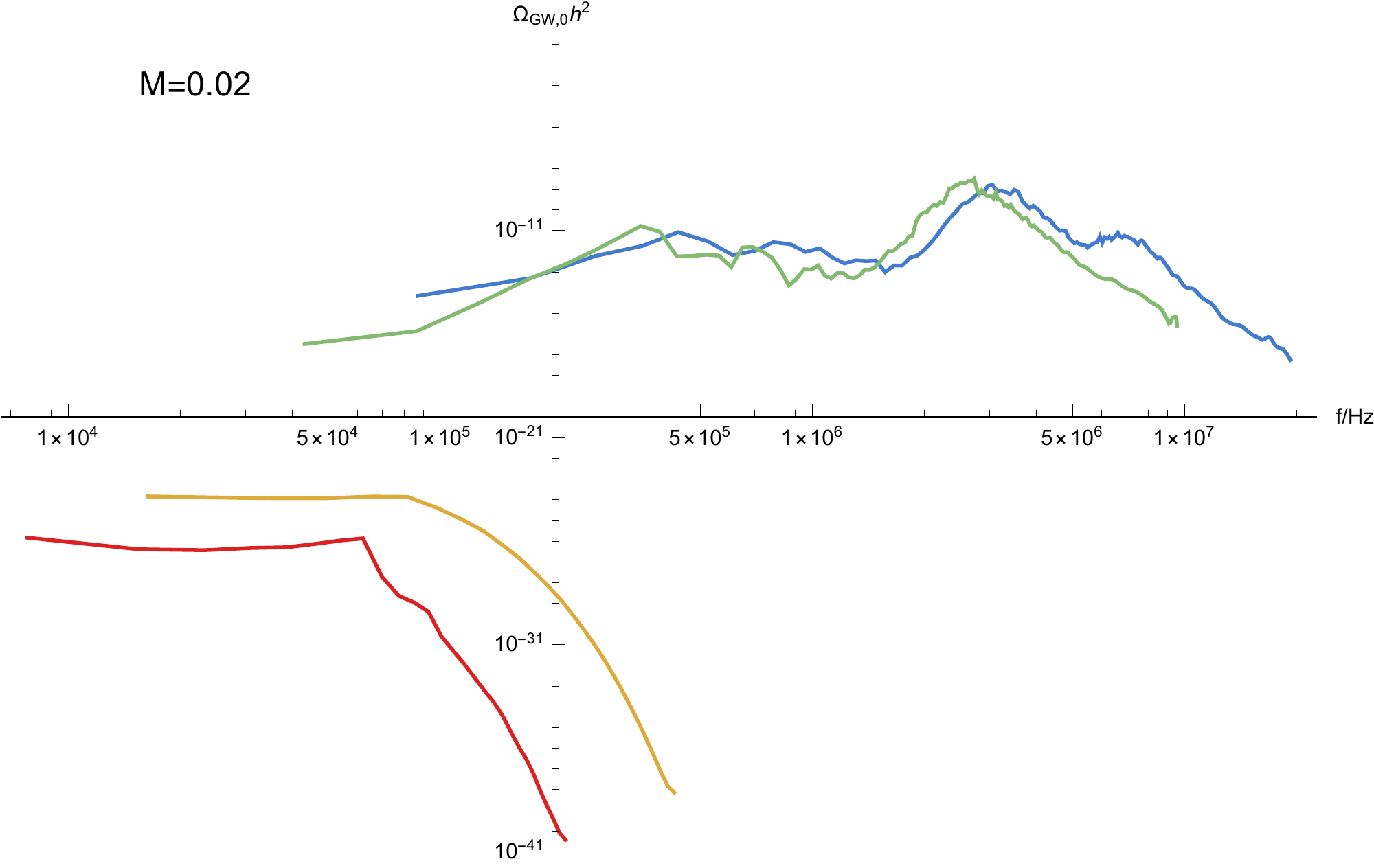}
	\raisebox{1\height}{\includegraphics[width=1.25in]{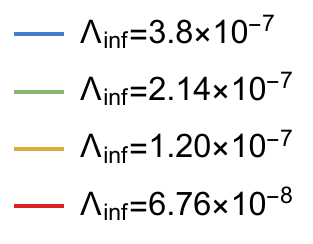}}
	\includegraphics[width=3in]{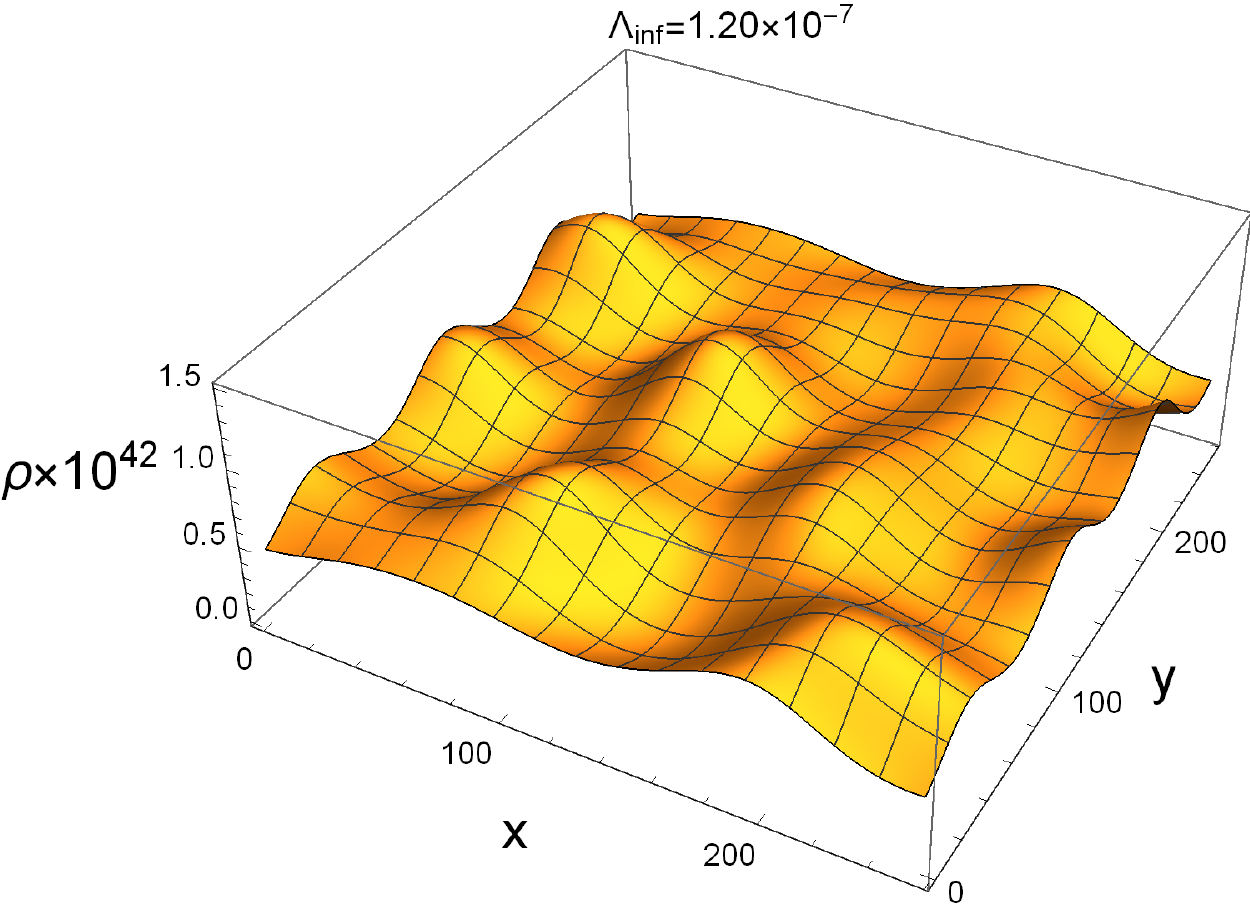}
	\includegraphics[width=3in]{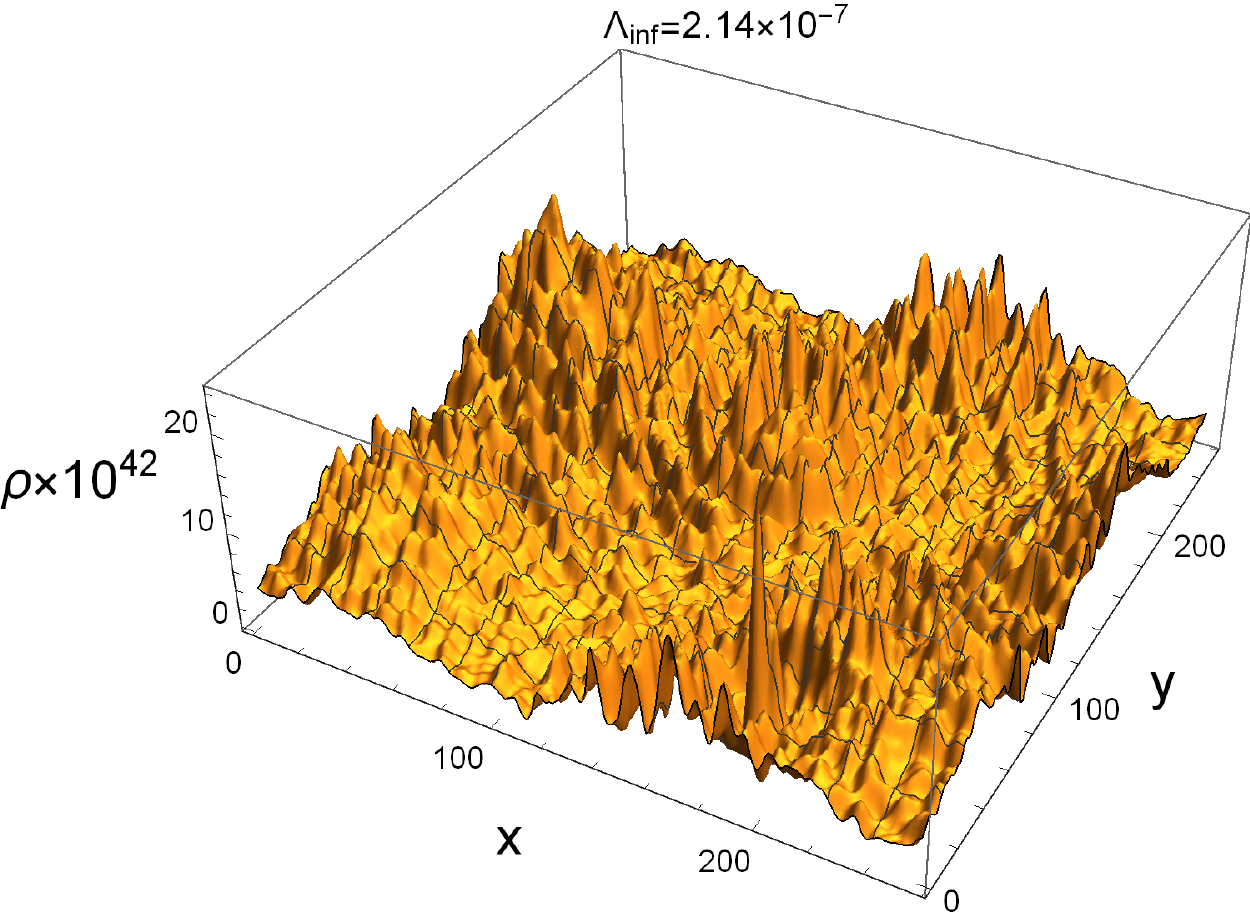}
	\caption{Top: the dependence of $\Omega_{\mathrm{GW},0}$ on $\Lambda_{\mathrm{inf}}$ for $M=0.02$. $\Omega_{\mathrm{GW},0}$ suddenly decreases for about $15$ orders of magnitude when $\Lambda_{\mathrm{inf}}$ decreases from $2.14\times 10^{-7}$ to $1.20\times 10^{-7}$. Bottom: the two-dimensional slices of energy density perturbations, where energy density perturbations are larger in $\Lambda_{\mathrm{inf}}=2.14\times 10^{-7}$ case.
	}
	\label{fig:lnu}
\end{figure*}

As shown in Fig.~\ref{fig:lnu}, the accurate value of $\Lambda_{c}$ is between $2.14\times 10^{-7}$ to $1.20\times 10^{-7}$ for $M=0.02$, where $\Omega_{\mathrm{GW},0}$ suddenly decreases when $\Lambda_{\mathrm{inf}}$ becomes slightly smaller than $\Lambda_{c}$. The small difference between the accurate value of $\Lambda_{c}$ and that obatined in the last section may come from the approximation in Eq.~\eqref{eq:EOMdelta}, where we have neglected the backreaction of $\delta\phi$ to the EOM and the decrease of the amplitude of coherent oscillation due to energy transfer. Taking into account the secondary effects yields more accurate $\Lambda_{c}$ but the treatment will become more complex. The profile of $\Omega_{\mathrm{GW},0}$ with $\Lambda_{\mathrm{inf}}=2.14\times 10^{-7}$ and $\Lambda_{\mathrm{inf}}=1.20\times 10^{-7}$ are different, where the former peaks at high-$k$ region while the latter exhibits a plateau in the low-$k$ region. Let us recall the evolution of $\Omega_{\mathrm{GW}}$ in Fig.~\ref{fig:evo}. The profile of $\Omega_{\mathrm{GW}}$ for $\Lambda_{\mathrm{inf}}=2.14\times 10^{-7}$ and $\Lambda_{\mathrm{inf}}=1.20\times 10^{-7}$ respectively correspond to $\Omega_{\mathrm{GW}}$ in the linear stage and fully nonlinear stage. This is shown more clearly by the two-dimensional slices of energy density perturbations in the lower panel of Fig.~\ref{fig:lnu}. The system remains in the linear stage for $\Lambda_{\mathrm{inf}}=1.20\times 10^{-7}$, while the nonlinear evolution is fully established for $\Lambda_{\mathrm{inf}}=2.14\times 10^{-7}$. 
By comparing the evolution of energy density perturbations and $\Omega_{\mathrm{GW}}$, we conclude that GWs are many orders of magnitude enhanced by nonlinear evolution. 
Fig.~\ref{fig:lnu} implies that once energy perturbations reach a certain threshold, the system will ultimately evolve into the fully nonlinear stage, even if the resonance becomes very weak. From Fig.~\ref{fig:lnu} one also finds that the length scale of perturbations becomes smaller in nonlinear evolution, corresponding to the increase of the GW peak frequency in the nonlinear stage.

In model (\ref{eq:Emodel}), the effective mass at the bottom of the potential is proportional to $M^{-1}$. As a result, $\bar{\phi}$ oscillates more rapidly as $M$ decreases and the energy stored in $\bar{\phi}$ tends to transfer into perturbations with smaller wavelengths. $\Omega_{\mathrm{GW},0}$ also decreases as $M^{2}$ as shown in Ref.~\cite{Lozanov:2019ylm}. Thus, on the one hand, $M$ needs to be small enough to trigger nonlinear evolution in low-scale inflationary models. On the other hand, decreasing $M$ makes $\Omega_{\mathrm{max}}$ smaller and difficult to detect.
For example, the most sensitive frequency of aLIGO is about $30\mathrm{Hz}$. We need $\Lambda_{c}< 10^{-11}$ to guarantee sufficient preheating takes place in the case the peak frequency is below $30\mathrm{Hz}$. Using the result in Fig.~\ref{fig:lin}, this condition gives an upper bound to the parameter $M\lesssim0.011$, and thus $\Omega_{\mathrm{max}}$ is also constrainted. 
According to the simulation results, $\Omega_{\mathrm{max}}\simeq 3\times 10^{-9}$ for $M=0.02$, and $\Omega_{\mathrm{max}}\simeq 1.5\times 10^{-8}$ for $M=0.04$, which is in consistant with the analytical result $\Omega_{\mathrm{max}}\propto M^{2}$ in Ref.~\cite{Lozanov:2019ylm}. Using this relation, $\Omega_{\mathrm{max}}$ is estimated as $7\times 10^{-10}$ for $M\simeq 0.01$, it is also the maximum of $\Omega_{\mathrm{max}}$ aLIGO could detect in model (\ref{eq:Emodel}).
In turn, detecting the peak frequency and the amplitude of GWs can help us  determine $\Lambda_{\mathrm{inf}}$ and $M$.


\section{Conclusion and Discussion}
\label{sec:con}
In this paper, we investigate the SGWB generated during preheating and the dependence of $\Omega_{\mathrm{GW}}$ on $\Lambda_{\mathrm{inf}}$.
We focus on a class of models where the effective potential has a quadratic minimum find $\Omega_{\mathrm{GW}}$ does not depend on $\Lambda_{\mathrm{inf}}$ only if $\Lambda_{\mathrm{inf}}$ is larger than a critical value $\Lambda_{\mathrm{c}}$. Since the initial value of $\delta\phi$ decreases as $\Lambda_{\mathrm{inf}}^{2}$, for $\Lambda_{\mathrm{inf}}<\Lambda_{c}$ the resonance is not strong enough to amplify $\delta\phi/\bar{\phi}$ to unity before the resonance ends, and the system stays in the linear stage. Inversely, for $\Lambda_{\mathrm{inf}}>\Lambda_{c}$ the resonance is strong enough, nonlinear evolution is fully established and $\Omega_{\mathrm{GW}}$ does not depend on $\Lambda_{\mathrm{inf}}$. 
We obtain $\Lambda_{c}$ in terms of $M$ in linear analysis, and confirm it later by the lattice simulations. Numerical results also show that, for $\Lambda_{\mathrm{inf}}$ slightly smaller than $\Lambda_{\mathrm{c}}$, $\Omega_{\mathrm{GW}}$ suddenly decreases more than ten orders of magnitude, and becomes challenging to be observed. We can find that $\Omega_{\mathrm{GW}}$ is immensely enhanced in nonlinear evolution by comparing the  energy density distribution in both $\Lambda_{\mathrm{inf}}\lesssim\Lambda_{c}$ and $\Lambda_{\mathrm{inf}}\gtrsim\Lambda_{c}$ cases. For $\Lambda_{\mathrm{inf}}\ll \Lambda_{\mathrm{c}}$, $\Omega_{\mathrm{GW}}$ is lower than $10^{-20}$ and proportional to $\Lambda_{\mathrm{inf}}^{8}$.

The peak frequency of GWs from the  preheating also provides useful information about $\Lambda_{\mathrm{inf}}$. As the trans-Planckian censorship conjecture suggests, $\Lambda_{\mathrm{inf}}$ should be lower than $10^{10}\mathrm{GeV}$. In this case, the peak frequency is lower than $10^{4}\mathrm{Hz}$, and GWs from the preheating are expected to be observed by interferometer observers such as LIGO, LISA, Taiji, DECIGO. Our work suggests that observing such GWs also gives constraints to the resonance strength and the model parameters. 

In preheating models where the potential is not quartic at the bottom, it is a common feature that the resonance strength decreases with time, and the analysis in this work is applicable to many preheating models. As shown in Ref.~\cite{Shtanov:1994ce,Kofman:1997yn,Lozanov:2017hjm}, with the expansion of the Universe the $k$-mode could stay in the resonance band only in the case $V(\phi)\propto \phi^{4}$ at the bottom. In other cases, for example, $V(\phi)\propto \phi^{2}$ or $\phi^{6}$ at the bottom, either the resonance bands disappears for $\phi\rightarrow 0$ or the modes are quickly redshifted out of the resonance bands. Hence, the resonance becomes weaker as the oscillation amplitude of $\bar{\phi}$ decreases, similar to the case considered in this work. 

As mentioned in Sec.~\ref{sec:nu}, $\Omega_{\mathrm{GW},0}$ decreases quadratically with $M$, and the value of $\Omega_{\mathrm{GW},0}$ LIGO could observe is less than about $7\times 10^{-10}$ because of the constraint $M<0.011$. For a smaller  $M$, it is more difficult to observe GWs from  the preheating. In comparison, for the models considered in our previous work, $\Omega_{\mathrm{max}}$ does not depend on model parameters. These models, inspired by string theory~\cite{McAllister:2008hb, Silverstein:2008sg}, have a characteristic cusp in the bottom, and GWs from the preheating in those models tends to be more strong.


\begin{acknowledgements}
This work is supported in part by the National Key Research and Development Program of China Grant No.2020YFC2201501, in part by the National Natural Science Foundation of China Grants No. 11435006, No.  11647601, No. 11690021, No. 11690022, No. 11821505, No. 11851302, No. 12047503, No. 11991052, No. 12075297, No. 12075298, No. 11947302, No. 12047559, in part by the China Postdoctoral Science Foundation under Grant
No.2020M680689, in part by the Strategic Priority Research Program of the Chinese Academy of Sciences Grant No. XDB23030100 and by Key Research Program of Frontier Sciences, CAS.
    
\end{acknowledgements}

\bibliography{infscale}

\end{document}